# Lateral scaling in carbon nanotube field-effect transistors

S. J. Wind,[*] J. Appenzeller, and Ph. Avouris
IBM T. J. Watson Research Center, P.O. Box 218, Yorktown Heights, NY 10598

We have fabricated carbon nanotube (CN) field-effect transistors with multiple, individually addressable gate segments. The devices exhibit markedly different transistor characteristics when switched using gate segments controlling the device interior versus those near the source and drain. We ascribe this difference to a change from Schottky barrier modulation at the contacts to bulk switching. We also find that the current through the bulk portion is independent of gate length for any gate voltage, offering direct evidence for ballistic transport in semiconducting CNs over at least a few hundred nanometers, even for relatively small carrier velocities.

Since the first demonstrations of carbon nanotube (CN) field-effect transistors (CNFETs),[1,2] steady improvements have been realized in their performance characteristics.[3,4,5,6] Along with these improvements come advancements in our understanding of how carbon nanotube devices work. It has recently been noted[7] that Schottky barriers are crucial for current transport in ambipolar CNFETs. It has also been found that, although most CNFETs reported to date appear to operate much the same as conventional silicon metal-oxide-semiconductor field-effect-transistors (MOSFETs), CNFETs are, in fact, Schottky-barrier (SB) MOSFETs.[8,9] This has important implications for a potential future CNFET-based technology in that CNFETs obey different scaling rules than do conventional MOSFETs.[10] In addition, the presence of Schottky barriers at the source and drain and the limited control demonstrated thus far with respect to the metallurgy at the metal-nanotube interface result in wide variability of the source and drain contact resistance. This has also made it very challenging to study the dependence of *intrinsic* CNFET characteristics on channel length. Since CNFET switching is dominated by these barriers,[11] and accordingly the sample resistance cannot in general be analyzed within the simple framework of a diffusive conductor, it is not straightforward to extract the actual transport properties of the nanotube itself. Attempts have been made to gain information about the mobility of carriers in nanotubes using channels of several micrometers in length.[12,13] The longer the tube, the more likely that transport is dominated by scattering inside the tube itself. However, for a quantitative analysis, the impact by the Schottky barriers must still be subtracted, resulting in a distinct uncertainty when analyzing the conductance.

In order to address this issue, we have fabricated CNFETs in a structure in which the gate field at the source and drain contacts can be decoupled from the gate field in the body of the nanotube. The device is shown schematically in Fig. 1(a). The gate is comprised of four independently addressable segments, separated from one another by a very short gap, ensuring that no part of the device is left ungated. This allows for modulation of the gate potential along the length of the device, and it enables operation of the device in several different electrical configurations, while maintaining the same physical and metallurgical properties. The device fabrication closely followed the process described in Ref. 3, with some minor variations in the gate lithography, where both high resolution and high placement accuracy were required for the patterning of the segmented gate. The scanning electron micrograph portion of the composite image in Fig. 1(a) shows a completed segmented gate overlapping the source and drain electrodes. The distance from source to drain is ~ 1 µm. The gate segments are ~ 480, 160, 80 and 240 nm, with ~ 20 - 25 nm gaps between segments. With a 20 nm-thick gate oxide, this narrow spacing ensures overlap of the gate field at the nanotube channel from neighboring segments.

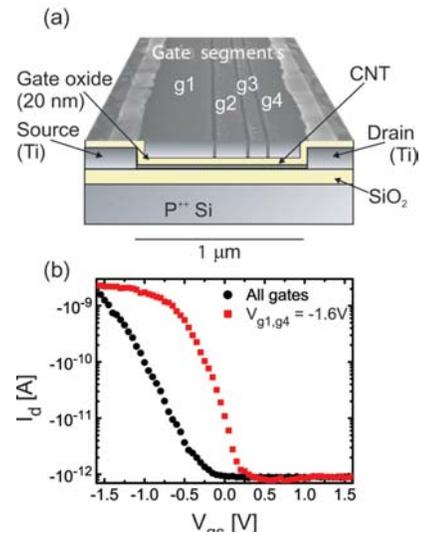

FIG. 1. (a) Schematic cross-section of segmented gate CNFET, overlayed by a scanning electron micrograph of the gate. (b) Transfer characteristics comparing subthreshold behavior for the case of all gate segments swept together versus sweeping only the interior gate segments, with the outer segments maintained at high potential.

Electrical measurements were performed in a variety of configurations, ranging from tying all four gate segments to the same potential and sweeping them as a single gate, to sweeping each segment individually, while the other segments were maintained at a fixed potential. Care was also taken to ensure that the gate segments were electrically isolated from one another. (The conductive substrate, used during the fabrication process to help identify working CNFETs, was held at ground potential.)

When the two outer segments, g1 and g4, are grounded, there is no conduction through the nanotube, independent of the potential on the inner segments. This is fully consistent with the case of an undoped carbon nanotube, and it is different from that of a doped CVD grown tube.[5] Since our CNFETs are enhancement mode transistors, i.e., normally off at zero gate voltage, the large gated segments close to source and drain prevent current flow for $V_{g1} = V_{g4} = 0$. As the potential on segments g1 and g4 is raised, the barriers in the source and drain region become thinner, and tunneling becomes more pronounced. In this way the transistor can be turned on provided that the inner gate segments are biased accordingly.

Figure 1(b) shows the CNFET drain current as a function of gate voltage for two different gate configurations: 1) all gate segments swept together from – 1.6 V to 1.6 V, and 2) only the inner two gate segments, g2 and g3, swept over the same range, while the outer segments, g1 and g4, are maintained at a fixed potential of – 1.6 V. The drain voltage is – 0.2 V. A clear difference can be seen in the inverse subthreshold slope, defined as $S = dV_{gs}/d(\log I_d)$. When all gate segments are operated as one, we find S ~ 400 mV/dec. However, when the device is operated by sweeping only the inner segments (with the outer segments held at – 1.6 V), the inverse subthreshold slope becomes significantly sharper, ~ 180 mV/dec. The change in behavior upon switching the inner segments versus using all the gate segments represents different modes of switching the CNFET, as illustrated in Fig. 2: When all gate segments are swept together, the main switching takes place as a result of modulation of the Schottky barriers at the source and drain;[7-9] raising the gate potential results in the thinning of these barriers, enabling increased tunneling. The properties of the gate dielectric (dielectric constant, thickness) determine the gate field at the contacts, and thus the subthreshold behavior.[10] On the other hand, when the outer segments are fixed at high bias, the barriers at the contacts are sufficiently transparent to allow relatively easy current injection. Sweeping the inner segments then bends the bands in the *bulk* of the CN, allowing thermal emission of carriers over the band edge, as in a conventional MOSFET. The result is a different subthreshold behavior, which is not limited by the properties of the gate dielectric. The portions of the CN covered by the outer gate segments can be viewed as an artificial source and drain, which are turned on by controlling the Schottky barriers at the metal/nanotube interface, similar to the straddle-gate MOSFET, in which a virtual source and drain are created in a silicon inversion layer by lateral engineering of the gate field or gate work function.[14]

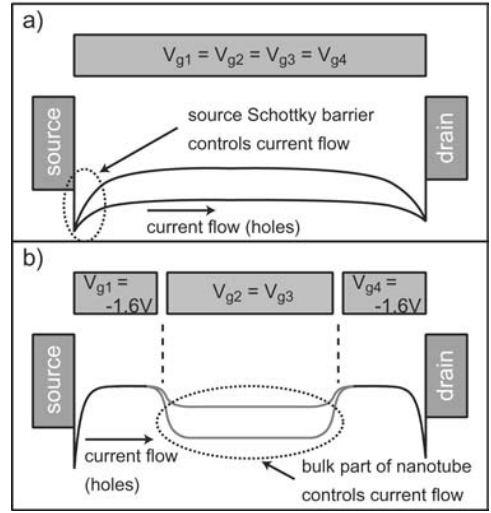

FIG. 2. Qualitative description of switching mechanisms represented by the two curves shown in Fig. 1 for two different gate voltages. (a) All gate segments tied together. (b) Inner gate segment potential is varied while the outer segments are biased at fixed potential.

We note that the measured value for the subthreshold slope is still far from the ideal value of 60 mV/dec. expected for thermally activated transport in a long channel device.[15] We believe that other factors, such as the presence of trapped charge near the oxide interface, may cause degradation of the subthreshold slope, as has been observed in other CNFETs.[16] This is also evident from the fact that the value of S observed here for the case of all gates swept together is somewhat high compared to other CNFETs with similar gate dielectric thickness.[3] The key point is the marked sharpening of the inverse subthreshold slope *within the same physical device structure* when the inner gate segments are used versus the outer segments. This sharpening cannot be accounted for by a change in the filling of interface traps, since their contribution to the subthreshold slope depends on the *ratio* of their capacitance, $C_{it}$, to the gate capacitance, $C_i$ ($S \sim C_{it}/C_i$).[15] Varying the gate length changes $C_{it}$ and $C_i$ simultaneously and is therefore not expected to affect S.

Evidence for bulk switching in CNFETs may have important technological implications. A review of published data on CNFET devices shows a wide variability in on-state current. This variability, in spite of the demonstrated ability to achieve high performance in CNFETs,[3,5,6] is a reflection of the *inability* to date to completely control the electrical properties of the metal/nanotube interface at the source and drain, and more specifically, the lineup of the valence band edge (for p-type CNFETs) relative to the Fermi level in the metal contact. A bulk switched transistor device is of great importance in this context.

We now turn our attention to the effects of varying the gate length within the bulk switching regime. Fig. 3 shows transfer characteristics for a CNFET with the two outer segments, g1 and g4, held at a fixed potential of – 1.6 V. The inner segments, g2 and g3, are swept between – 1.6 V and 1.6 V in two different configurations: 1) Both g2 and g3 swept together and 2) g2 swept individually, with g3 held at – 1.6 V.[17] The drain voltage is – 0.2 V. These two configurations

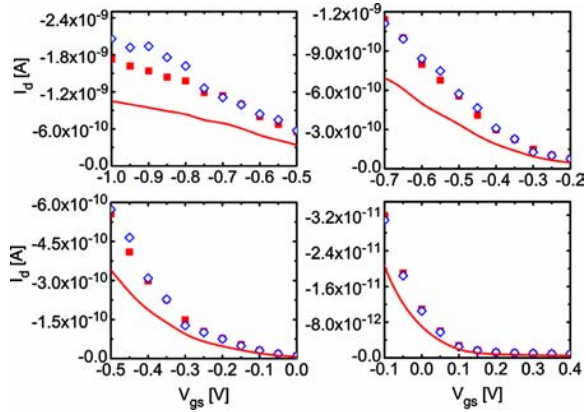

FIG. 3. Drain current vs. gate voltage for two different effective gate lengths, achieved by sweeping both segments g2 and g3, where $L_g \approx 260$ nm (solid symbols) vs. sweeping only g2, where $L_g \approx 160$ nm (open symbols). The solid line indicates the *expected* drain current for a 260 nm channel assuming diffusive transport, calculated by scaling the data for segment g2 by a factor of 1.625.

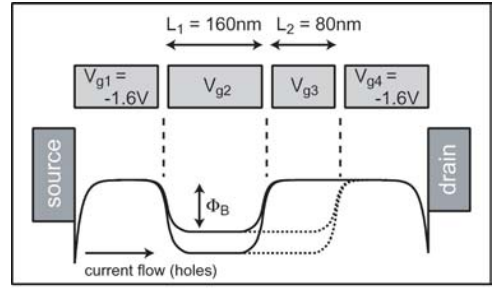

FIG. 4. Qualitative description of the energy band situation in the case of bulk mode operation for two different gate lengths and two different gate voltages

correspond to different effective gate lengths: ~ 260 and 160 nm, respectively. Note, that these configurations eliminate the impact of the field-dependent Schottky barriers in the source and drain region. Fig. 3 shows the drain current on a linear scale for four different ranges of gate voltage. In the presence of backscattering (i.e., "diffusive" transport),[18] the current is expected to vary as ~ $\mu/L_g$, where $\mu$ is the carrier mobility, and $L_g$ is the gate length. Thus, the current should decrease linearly with increasing gate length, as indicated by the solid curves in Fig 3. (The solid curves are calculated by scaling the current for segment g2 by a factor of 1.625, the ratio of the two gate lengths shown.) No such variation is seen for the two gate lengths probed, indicating that there is no apparent back-scattering occurring inside the nanotube device for the gate lengths considered here. Interestingly, this statement is true for any given gate voltage, i.e., any band bending situation inside the carbon nanotube. This behavior is illustrated in Fig. 4, where the effects of the potential from the inner gate segments can be seen for two gate voltages in the two different gate configurations discussed above. Carriers entering from the virtual source region under segment g1 can traverse the regions under g2 and g3 if they have enough energy to overcome the gate modulated barriers. In the case of a diffusive channel, scattering in these regions would result in a lower current for the longer channel. The fact that the observed current for the two different channel lengths is the same within the accuracy of our measurement is a clear proof that transport through these regions is ballistic. Even more importantly, no scattering is observable even for the smallest drain currents measured. Since a reduction of current goes along with a higher potential barrier, which implies an increasingly smaller hole velocity in the gated nanotube segment, our experiment provides first evidence that coulomb scattering is suppressed even for the smallest velocities considered here. With respect to device applications, this observation implies that dc characteristics cannot be improved by reducing the gate length of a CNFET, in contrast to a conventional MOSFET. However, it is expected that ac characteristics should be affected, since the time-of-flight through the nanotube channel changes with gate length. Further experiments are necessary to clarify this point.

Appenzeller et al.[8] have already pointed out the lack of any length dependence of certain CNFET parameters. That observation highlighted the dominance of Schottky barriers on CNFET performance. Here we make a more far-reaching statement about transport inside a semiconducting carbon nanotube. While in our former work, information regarding scattering in a CNFET could only be indirectly inferred, our present experiment enables us to directly probe the intrinsic carbon nanotube properties. Since even a four-probe measurement on a carbon nanotube perturbs the transport conditions inside the tube,[19] and Schottky barriers prevent a simple analysis of two-terminal transport data, our approach is the most direct way to obtain information about scattering in a CNFET.

In conclusion, by fabricating CNFETs in a segmented gate structure, we have been able to probe their properties as a function of gate length. We find no length dependence of the on-state and off-state, consistent with ballistic transport over distances of a few hundred nanometers, i.e., lengths relevant for field-effect devices. In addition, we have been able to clearly distinguish between Schottky barrier switching and bulk switching in CNFETs, which suggests paths which may be pursued for future nanoelectronic applications.

The authors thank P. M. Solomon for insightful and stimulating discussions, and B. Ek and R. Sicina for thin film deposition.

[*] Present address: Department of Applied Physics and Applied Mathematics, Columbia University, 1020 Shapiro CEPSR, 530 W. 120th St., New York, NY 10027